\documentstyle[prl,aps]{revtex}
\input{epsf.tex}
\input psfig
\twocolumn
\draft
\include{psfig}
\begin{document}
\twocolumn[\hsize\textwidth\columnwidth\hsize\csname %
@twocolumnfalse\endcsname

\title{Evolution of the $\nu=1$ Ground-State in Coupled Double Quantum
  Wells: Optical Evidence for Broken-Symmetry States}

\author{Michael J. Manfra$^{1,*}$, Justin C. Pniower$^1$, Bennett B.
  Goldberg$^1$, \\ Aron Pinczuk$^2$, Vittorio Pellegrini$^3$, Loren N.
  Pfeiffer$^2$, and Ken W. West$^2$}

\address{$^1$Department of Physics, Boston University, Boston,
  Massachusetts 02215}

\address{$^2$Bell Laboratories, Lucent Technologies, Murray Hill, New Jersey 07974}

\address{$^3$Scuola Normale Superiore and INFM, Piazza dei
  Cavalieri 7, I-56126, Pisa, Italy}

\date{\today}
\maketitle
\begin{abstract}  We present the first magneto-absorption studies of coupled electron
  double layers in the quantum Hall regime.  Optical absorption spectra
  in the vicinity of total filling factor $\nu=1$ reveal intriguing
  behavior that have no analog in the single electron layer $\nu=1$
  state and demonstrate the interplay between single-particle tunneling
  and inter-layer Coulomb effects.  The spectra provide direct evidence
  of a ground-state that evolves from a region dominated by
  single-particle tunneling to a regime in which inter-layer Coulomb
  interactions determine the nature of the ground-state.  Moreover the
  spectra provide the first direct evidence that the {\it
    incompressible} ground-state at $\nu=1$ in the many-body regime is
  not fully pseudospin polarized and is sensitive to the effects of
  quantum fluctuations in the pseudospin variable.

\bigskip \bigskip
\noindent
\pacs{PACS numbers:  73.20.Dx, 73.20Mf, 78.30.Fs}
\end{abstract}
\twocolumn
\vspace{2mm}
]

The fractional quantum Hall effect (FQHE) is understood as arising from
a gap in the excitation spectrum of a two-dimensional (2D) electron
system caused by strong Coulomb interactions in the presence of a large
perpendicular magnetic field sufficient to quench the electronic kinetic
energy.  Generally, such quantum Hall states can be represented by a
single component many-body wavefunction describing the correlated 2D
orbital motion of the electrons \cite{Laughlin}.  An interesting twist
to the problem is added when some internal degree of freedom is not
frozen out by the magnetic field and persists as a dynamical variable.
The spin degree of freedom in the limit of small Zeeman coupling is a
prime example of such a {\it multicomponent} quantum Hall system.  The
impact of spin for determining the excitation spectrum around filling
factor $\nu=1$ in the single-layer 2D electron system has been an area
of recent intense theoretical and experimental inquiry
\cite{sondhi-93-1,Fertig-94-1,barrett-95-1,schmeller-95-1,goldberg-96-3,maude2,manfra-96}.
There now exists a large body of evidence that the quantum Hall state at
$\nu=1$ in the single-layer system is more appropriately viewed as a
``fractional'' state inasmuch as the gap in its excitation spectrum
survives the collapse of the single-particle spin gap.

Another multicomponent quantum Hall system occurs in the coupled double
quantum well (DQW) structure \cite{Boebinger}. By growing two 2D
electron layers in close proximity, a new degree of freedom associated
with the layer index is introduced.  In direct analogy with the spin-1/2
system, the layer index is associated with a double-valued pseudospin
variable.  An electron in the ``upper'' layer is in an eigenstate of the
pseudospin operator $S_z$ with eigenvalue +1.  Similarly an electron in
the ``lower'' layer has eigenvalue -1.  In the presence of tunneling,
symmetric and anti-symmetric combinations of the eigenstates of $S_z$
can be constructed which are eigenstates of $S_x$.  At B=0, these states
are separated by a single-particle tunneling energy gap, $\Delta_{SAS}$.
At total filling factor $\nu=1$, the non-interacting ground-state will
consist of a fully populated symmetric state of the spin-up branch of
the lowest Landau level (LLL).  In the pseudospin picture, this state is
fully pseudospin polarized along the $\hat x$ direction.  Nevertheless,
theory has anticipated that inter-layer Coulomb interactions will
profoundly alter the nature of the quantum Hall states of the DQW at
$\nu=1$ \cite{Mac,KunYang,moon-prb95,Perspectives}.  Indeed, the
reduction of symmetry introduced by the inter-layer Coulomb interaction
is expected to introduce quantum fluctuations which destroy the full
pseudospin polarization of the $\nu=1$ ground-state \cite{Perspectives}.
Despite a fluctuating pseudospin polarization, the system remains
incompressible and exhibits well-defined quantum Hall state.  A non
fully pseudospin polarized ground-state concomitant with an excitation
gap represents one of the most unusual and non-trivial aspects of
inter-layer coherence.

To date, most experimental investigations of coupled double layer
quantum Hall systems have been limited to transport studies
\cite{Boebinger,Murphy,Lay}.  At total filling factor $\nu=1$, the
appearance of incompressible quantum Hall states or compressible
ground-states is determined by the delicate balancing of the tunneling
gap $\Delta_{SAS}$, the inter-layer Coulomb energy scale set by the
distance $d$ between 2D electron layers, and the intra-layer Coulomb
correlations determined by $l_0$, the magnetic length, where $l_0 =
(\hbar / eB)^{1\over2}$.  In seminal work, Murphy {\it et al.}
\cite{Murphy} constructed a phase diagram for $\nu=1$ in the DQW
structure. In addition to determining a well-defined phase boundary
between regimes which support a gapped $\nu=1$ state and those for
which the ground-state at $\nu=1$ is compressible, they also found
that the double layer $\nu=1$ quantum Hall state evolves continuously
from a regime in which the gap is largely determined by
single-particle tunneling to a regime where the gap is necessarily of
a many-body origin.  Their measurements in the weak tunneling regime
suggest that the $\nu=1$ quantum Hall state in the DQW structure also
survives the collapse of the single-particle gap.  While it is clear
that the ground-state must evolve as the tunneling strength is reduced
relative to the inter-layer Coulomb interactions, transport cannot
clearly distinguish between the two regimes nor directly probe the
pseudospin configuration of the ground-state, leaving many intriguing
questions open.

In this letter, we present to our knowledge the first magneto-absorption
measurements of the coupled DQW system in the quantum Hall regime.  Our
discussion will focus on total filling factor $\nu=1$.  We have studied
a number samples in order to investigate the optical response of the
coupled 2D electron system as the sample parameters are tuned from a
regime where a gap in the single-particle spectrum accounts for the
quantum Hall effect at $\nu=1$ to a regime where a quantized Hall state
at $\nu=1$ reflects a correlated many-body ground-state.  The observed
spectra provide direct evidence that the DQW system in the many-body
regime exactly at $\nu=1$ is {\it not fully pseudospin polarized}.  It
is important to note that all samples used in this study exhibit
well-defined quantum Hall plateaux and longitudinal resistivity minima
at $\nu=1$.  While transport shows similar behavior, interestingly, the
observed optical spectra display qualitatively different behavior
depending on the sample's position on the $\nu=1$ phase diagram.  We
suggest that the observed spectral changes reflect an evolution of the
ground-state at $\nu=1$ driven by quantum fluctuations in the pseudospin
degree of freedom.

Optical probes provide complimentary means for probing the integral and
fractional quantum Hall regimes, but their application to the coupled
double electron layer system at $\nu=1$ has been rather limited.  An
optical emission study \cite{Vitto3} of the DQW structure has observed
anomalies at $\nu=1$ which have been associated with a change in the
pseudospin state, but a complete understanding of this emission data is
still lacking.  Recently, inelastic light scattering has been
successfully employed to observe a collapse of collective spin-wave
excitations at {\it even} filling factors in the DQW system
\cite{Vito1,Vito2}.  Magneto-absorption spectroscopy relies on its
ability to discriminate between occupied and unoccupied states in the
vicinity of the Fermi level.  Absorption can only occur into {\it
  unoccupied} states and therefore monitoring absorption into the
symmetric and antisymmetric levels as the Fermi level sweeps through
$\nu=1$ may be used to elucidate the ground-state pseudospin
configuration.

We discuss in detail absorption spectra obtained from two distinct
modulation-doped DQW samples grown by molecular beam epitaxy.  Sample A
consists of two identical 180$\AA$ GaAs quantum wells separated by a
79$\AA$ Al$_{0.1}$Ga$_{0.9}$As undoped barrier layer.  The electron
density in this sample is $n=6.3 \times 10^{10}$ cm$^{-2}$ and the
mobility is close to $10^6$cm$^2$/Vs at low temperatures.  At $B=0$,
$\Delta_{SAS}=0.7$meV in this sample, as measured by inelastic light
scattering \cite{Annette}.  Sample B consists of two identical 180$\AA$
GaAs quantum wells separated by a 31$\AA$ undoped Al$_{0.3}$Ga$_{0.7}$As
barrier.  The density is $n=1.3 \times 10^{11}$cm$^{-2}$ with a mobility
of $10^6$cm$^2$/Vs.  The tunneling gap $\Delta_{SAS}$ for sample B is
$0.4$meV.  In order to perform transmission studies, the samples are
mounted strain-free on Corning glass which has a coefficient of thermal
expansion matched to GaAs.  The bulk substrate is then removed via a
combination of mechanical polishing and a modified chemical jet-etching
process \cite{LePore}.
\begin{figure}[h]
\begin{center}
  \mbox{\psfig{figure=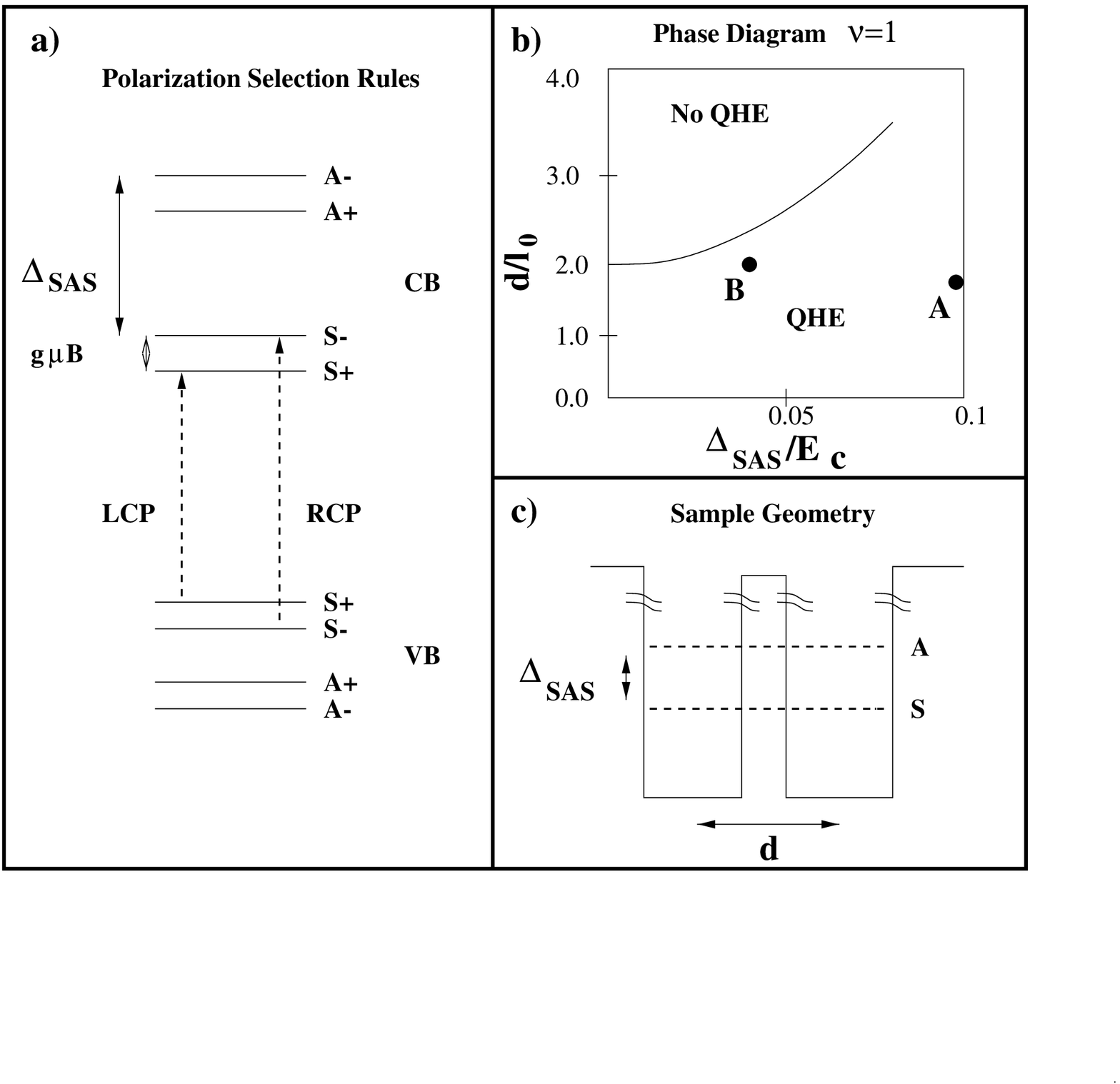,width=3.5in}} \end{center}
\caption{(a) Schematic representation of polarization selection rules
  governing inter-band absorption at $\nu=1$.  The dashed lines
  correspond to the two lowest-energy transitions in the left-circular
  polarization (LCP) and the right-circular polarization (RCP).  (b)
  Phase diagram of $\nu=1$ in the DQW structure constructed by Murphy
  {\it et al.}  \protect \cite{Murphy}.  The $x$ axis measures the
  tunneling gap $\Delta_{SAS}$ in units of the basic Coulomb energy
  $E_c=e^2/\epsilon l_0$.  The $y$ axis scale, $d/l_0$, measures the
  ratio of intra-layer to inter-layer Coulomb interactions.  $d$ is the
  center-to-center well separation and $l_0$ is the magnetic length.
  (c) Schematic of DQW structure used in these studies with the
  symmetric-antisymmetric gap, $\Delta_{SAS}$,
  displayed.\label{fig:phase.eps}}
\end{figure} 
\noindent
Accessible temperatures were in the range of $0.5K \le T \le 4.2K$.
The measured absorption coefficient is given as $\alpha = -1/L_w
ln[I(B)/I(0)]$, where $L_w$ is the quantum well width, $I$ the
measured transmission intensity, and $B$ the magnetic field.

\begin{figure}[h]
\begin{center}
  \mbox{\psfig{figure=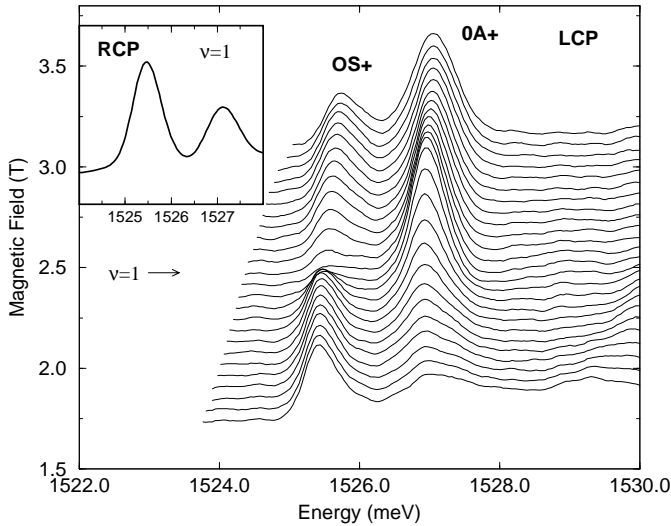,width=3.5in}} \end{center}
\caption{Low-energy absorption in LCP at T=0.53K of sample A.  The
  transition labeled 0S+ monitors absorption into the symmetric-state of
  the LLL while 0A+ corresponds to absorption into the antisymmetric
  state.  Note the strong quenching of the lowest energy transition
  exactly at $\nu=1$.  The inset displays the absorption in RCP at
  $\nu=1$.  The significantly different behavior observed in RCP
  indicates that the selection rules are active. \label{fig:fig1.eps}}
\end{figure} 
\noindent

Figure \ref{fig:phase.eps} displays the relevant inter-band transitions
and polarization selection rules in the vicinity of $\nu=1$.  Also shown
is a reproduction of the phase diagram constructed by Murphy {\it et
  al.}  \cite{Murphy} and discussed in the introduction.  Optical
selection rules \cite{goldberg-96-3,manfra-96} dictate that the
lowest-energy transition in the left-circular polarization (LCP) of the
light field will monitor absorption into the symmetric spin-up state of
the LLL around $\nu=1$.  Figure \ref{fig:fig1.eps}
displays the LCP spectra in the vicinity of $\nu=1$ for sample A at a
temperature T=0.53K.  The inset displays the absorption at $\nu=1$ in
RCP and highlights the distinctly different behavior observed in the two
polarizations. The ratio of 20:1 in absorption between RCP and LCP at
1525.5 meV indicates that the optical selection rules shown in Fig.
\ref{fig:phase.eps} are active.  A tunneling gap of
$\Delta_{SAS}=0.7$meV puts sample A into a regime where the gap at
$\nu=1$ should be largely a single-particle effect and the ground-state
at $\nu=1$ should be nearly fully pseudospin polarized along $\hat x$,
i.e. the ground-state at $\nu=1$ consists of a fully occupied symmetric
state of the LLL.

\begin{figure}[h]
\begin{center}
  \mbox{\psfig{figure=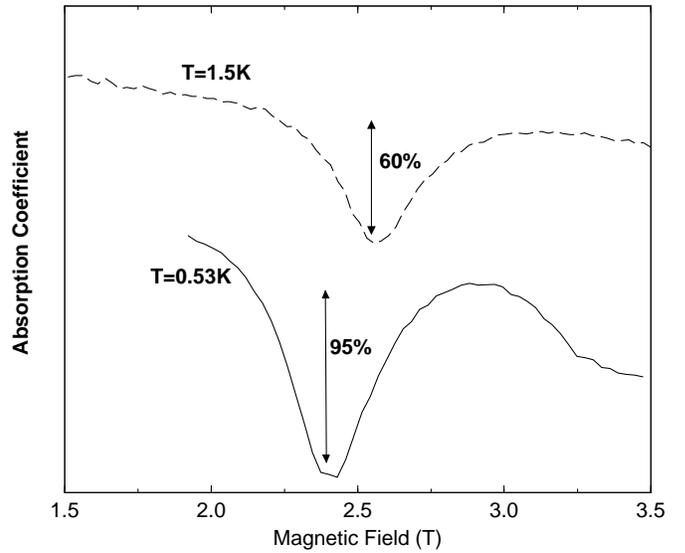,width=3.5in}} \end{center}
\caption{Quenching of absorption into the symmetric state of the LLL as
  the Fermi level moves through $\nu=1$ at T=1.5K and T=0.53K for sample
  A.  Note that the quenching is well developed at T=1.5K.  The
  quenching at $\nu=1$ is measured relative to the absorption level at
  2T and 3T. \label{fig:6-15-94-lcp.eps}}
\end{figure} 
\noindent

In the region of $\nu=1$ we observe two low-energy transitions whose
final states we assign to the symmetric spin-up $(0S+)$ and
antisymmetric spin-up $(0A+)$ states of the LLL.  The most striking
feature in Fig.\ref{fig:fig1.eps} is the strong quenching of the
lowest-energy transition in LCP as the Fermi level passes through
$\nu=1$.  This behavior is very reminiscent of the quenching seen in the
single-layer system at $\nu=1$ where the ground-state is fully {\it
  spin} polarized and skyrmions are present for small deviations from
$\nu=1$ \cite{goldberg-96-3,manfra-96}. The quenching of the
lowest-energy LCP transition is an optical {\it signature} of a
ferromagnetically aligned ground-state in the single-layer system.  The
observed minimum of the absorption in the DQW structure shows that the
lowest-energy transition in LCP is sensitive to the ground-state
occupation of the symmetric level at $\nu=1$.  Figure
\ref{fig:6-15-94-lcp.eps} displays the magnetic field dependence of this
absorption minimum as the Fermi level moves through $\nu=1$ at two
different temperatures.  It is clear that the absorption minimum is
already well-developed by T=1.5K and that below T=1.5K there are
no qualitative changes in the absorption.  All small intensity
variations of the low-energy transitions have a monotonic temperature
dependence indicating activated behavior.  This is strong indication
that there are no unoccupied symmetric states of the LLL exactly at
$\nu=1$ at zero temperature.  Indeed, the observed spectral features are
consistent with the expectation of a fully pseudospin polarized
ground-state at $\nu=1$ for sample A in the regime of relatively strong
tunneling.

\begin{figure}[t]
\begin{center}
  \vbox{\psfig{figure=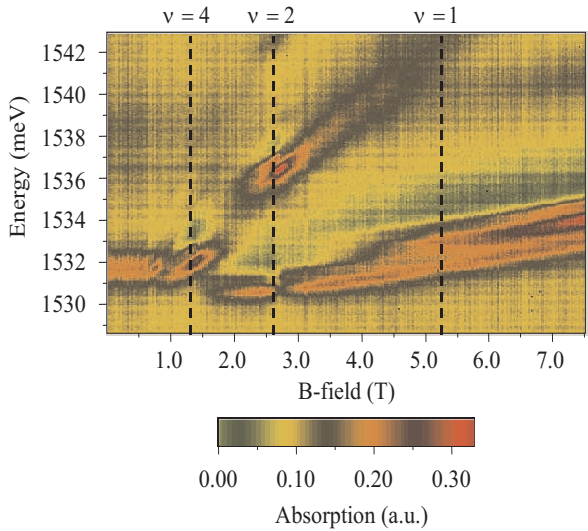,angle=-90,height=3in,width=3.5in}} \end{center}
\caption{Intensity plot of the absorption in LCP for sample B
at T=1.5K.  The positions of several filling factors are displayed.
Note, in sharp contrast to sample A, the lack of quenching of the
lowest-energy transition at $\nu=1$.\label{fig:15.eps}}
\end{figure}

Qualitatively different behavior is observed in sample B.  Sample B
possesses a much smaller single-particle tunneling gap,
$\Delta_{SAS}=0.4$meV.  Figure \ref{fig:15.eps} shows an intensity map
of the absorption for sample B in LCP at T=1.5K.  At low fields $(B\le
1.3T$, and $\nu \ge 4)$ the absorption has excitonic character.
Nevertheless, even in the low field regime the absorption clearly
responds the the presence of the incompressible quantum Hall states of
the 2D electron system.  Intensity maxima and/or minima are present at
filling factors $\nu=8$, 6, 4 and 2.  In addition to showing optical
sensitivity to the position of the Fermi level, the optical anomalies
allow for an accurate determination of the electron density.  As in
sample A, the spectra in sample B show strong polarization dependence.
At fields corresponding to $\nu \le 3$, the absorption develops a linear
field dependence indicative of well-developed Landau level-like
transitions.

\noindent

For sample B in the region of $\nu=1$, the response of the lowest-energy
transition in LCP is strikingly different than that seen in sample A.
Rather than observing a quenching of the absorption, the lowest-energy
LCP transition displays a weak maximum as the Fermi level is swept
through $\nu=1$ at T=0.5K.  The observed lack of an optical anomaly in sample B
implies the existence of a finite density of available states in the
symmetric state of the LLL at $\nu=1$.  The observation of non-zero
absorption persists to the lowest accessible temperature of T=0.5K.
These observations suggest that the ground-state at $\nu=1$ in sample B
is not fully pseudospin polarized along the $\hat x$ direction.

\begin{figure}[h]
\begin{center}
  \mbox{\psfig{figure=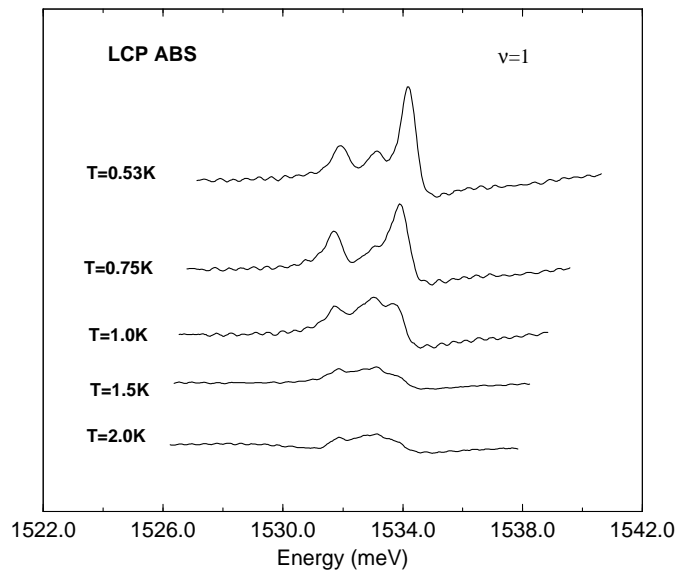,height=3in,width=3.5in}} \end{center}
\caption{Individual absorption spectra in LCP at $\nu=1$ for sample B taken at
  various temperatures.  The spectra change dramatically between 2K and
  1K, suggesting a rapidly evolving ground-state.  Below T=0.8K the
  lowest-energy transitions appear activated.
  \label{fig:tempabslcp.eps}}
\end{figure}

Another indication of a drastically different ground-state in sample B
comes from an examination of the temperature dependence of the
absorption at $\nu=1$.  Figure \ref{fig:tempabslcp.eps} displays
individual spectra at $\nu=1$ taken at various temperatures.  In sharp
contrast to sample A, the absorption in sample B changes dramatically as
the temperature is reduced from T=2K to T=1K.  Interestingly, the
absorption into the lowest-energy LCP transition actually {\it
  increases} at low temperatures, indicating that our observations are
not limited by the thermal population of excited states.  Below T=0.8K
the spectral changes stabilize and indicate activated behavior. Such
unusual temperature dependence has been observed in transport as a
deviation from simple activated behavior at relatively low temperatures
$(T \leq 0.5K)$, despite the presence of a measured gap that exceeds this
temperature by a factor of 20 \cite{Murphy}.  Theoretically this
behavior has been associated with a thermally induced collapse of the order
that produced the collective gap \cite{Perspectives}.

Incomplete pseudospin polarization of the ground-state at $\nu=1$ in
these systems is due to the reduction in symmetry from $SU(2)$ to $U(1)$
caused by the inter-layer Coulomb interactions and has been anticipated
theoretically \cite{Mac,KunYang,moon-prb95,Perspectives}.  The
symmetry-breaking term of the inter-layer Coulomb interaction does not
commute with the total pseudospin operator, $[V_{sb},S] \ne 0$,
introducing quantum fluctuations, and making total pseudospin no longer
a sharp quantum number \cite{Perspectives}.

It is important to note that the incompletely pseudospin polarized state
remains incompressible and exhibits a gap for both neutral and charged
elementary excitations.  As a consequence, a well-defined $\nu=1$
quantum Hall state is observed in transport in sample B.  For a
non-interacting model, partial pseudospin polarization would be
incompatible with a quantum Hall plateau. Thus the observed spectra
represent a non-trivial manifestation of inter-layer coherence.

In conclusion, we have presented magneto-absorption data from the
coupled DQW structure in the quantum Hall regime around $\nu=1$ which
reveal intriguing optical anomalies that have no analog in the
single-layer system.  Magneto-absorption appears to be sensitive to
changes in the ground-state of the electron system in a manner
inaccessible to transport.  Finally, low-temperature spectra obtained
from a sample with a relatively small tunneling gap,
$\Delta_{SAS}=0.4$meV, suggest that this ground-state at $\nu=1$ is not
fully pseudospin polarized.  The observed absorption is consistent with
the presence of quantum fluctuations caused by the reduction in the
symmetry of the Coulomb interactions.  Further optical experiments are
underway to explore more of this rich phase diagram.

We acknowledge valuable conversations with Allan MacDonald, David
Broido, and Luis Brey.  The work completed at Boston University was
supported by the NSF under grant number DMR 9701958.
\vspace{0.2in} \\
$^*$Present address: Bell Laboratories, Lucent Technologies, Murray
Hill, New Jersey, 07974

\bibliography{mike_ref}

\begin{thebibliography}{10}

\bibitem{Laughlin}
R.~B. Laughlin, Phys. Rev. Lett. {\bf 50},  1395  (1983).

\bibitem{sondhi-93-1}
S.~L. Sondhi, A. Karlhede, S.~A. Kivelson, and E.~H. Rezayi, Phys. Rev. B {\bf
  47},  16419  (1993).

\bibitem{Fertig-94-1}
H.~A. Fertig, L. Brey, R. Cote, and A.~H. MacDonald, Phys. Rev. B {\bf 50},
  11018  (1994).

\bibitem{barrett-95-1}
S.~E. Barrett {\it et~al.}, Phys. Rev. Lett. {\bf 74},  5112  (1995).

\bibitem{schmeller-95-1}
A. Schmeller, J.~P. Eisenstein, L.~N. Pfeiffer, and K.~W. West, Phys. Rev.
  Lett. {\bf 75},  4290  (1995).

\bibitem{goldberg-96-3}
E.~H. Aifer, B. Goldberg, and D. Broido, Phys. Rev. Lett. {\bf 76},  680
  (1996).

\bibitem{maude2}
D.~K. Maude {\it et~al.}, Phys. Rev. Lett. {\bf 77},  4604  (1996).

\bibitem{manfra-96}
M. Manfra {\it et~al.}, Phys. Rev. B {\bf 54},  R17327  (1996).

\bibitem{Boebinger}
G. Boebinger, H. Jiang, L. Pfeiffer, and K. West, Phys. Rev. Lett. {\bf 64},
  1793  (1990).

\bibitem{Mac}
A.~H. MacDonald, P.~M. Platzman, and G.~S. Boebinger, Phys. Rev. Lett. {\bf
  65},  775  (1990).

\bibitem{KunYang}
K. Yang, Phys. Rev. Lett. {\bf 72},  732  (1994).

\bibitem{moon-prb95}
K. Moon {\it et~al.}, Phy. Rev. B. {\bf 51},  5138  (1995).

\bibitem{Perspectives}
{\em Perspectives in Quantum Hall Effects}, edited by S.~D. Sarma and A.
  Pinczuk (Wiley Interscince, New York, 1997), chp. 5.

\bibitem{Murphy}
S.~Q. Murphy, J. Eisenstein, L. Pfeiffer, and K. West, Phys. Rev. Lett. {\bf
  72},  728  (1994).

\bibitem{Lay}
T. Lay {\it et~al.}, Phys. Rev. B {\bf 50},  17725  (1994).

\bibitem{Vitto3}
V. Pellegrini {\it et~al.},  in {\em High Magnetic Fields in the Physics of
  Semiconductors II}, edited by G. Landwehr and W. Ossau (World Scientific,
  Singapore, 1997), Vol.~2, p.\ 681.

\bibitem{Vito1}
V. Pellegrini {\it et~al.}, Phys. Rev. Lett. {\bf 78},  310  (1997).

\bibitem{Vito2}
V. Pellegrini {\it et~al.}, Science {\bf 281},  799  (1998).

\bibitem{Annette}
A.~S. Plaut {\it et~al.}, Phys. Rev. B {\bf 55},  9282  (1997).

\bibitem{LePore}
J.~J. LePore, Journal of Applied Physics {\bf 51},  6441  (1980).

\end{thebibliography}
\bibliographystyle{prsty}

\end{document}